\documentclass[conference]{IEEEtran}
\IEEEoverridecommandlockouts
\usepackage{cite}

\ifCLASSINFOpdf
\else
\fi

\usepackage[cmex10]{amsmath}
\usepackage{mathtools}
\usepackage{amssymb}
\usepackage{array}
\usepackage[pdftex]{graphicx}
\usepackage{multirow}
\usepackage{cases}
\usepackage[tight,footnotesize]{subfigure}
\usepackage{algorithm} %format of the algorithm
\usepackage{algorithmic} %format of the algorithm
 \usepackage[cmex10]{amsmath}
\usepackage{amssymb}
\usepackage{booktabs}
\usepackage{url}
\usepackage{graphicx}
\usepackage{subfigure}
\newtheorem{assumption}{Assumption}

\newtheorem{theorem}{Theorem}

\usepackage{bbm}
\newcolumntype{C}[1]{>{\vspace{0.5em}\begin{minipage}{#1}\centering\let\newline\\
\arraybackslash\hspace{0pt}}m{#1}<{\end{minipage}\vspace{0.5em}}}

\begin{document}

%\fontsize{26pt}{26pt}\selectfont{
\title{Incentivizing Users of Data Centers \\Participate in The Demand Response Programs\\
via Time-Varying Monetary Rewards}

\author{\IEEEauthorblockN{Yong Zhan\IEEEauthorrefmark{1}, Du Xu\IEEEauthorrefmark{1}, Hongfang Yu\IEEEauthorrefmark{1}, Shui Yu\IEEEauthorrefmark{4}}

\IEEEauthorblockA{\IEEEauthorrefmark{1}Key Laboratory of Optical Fiber Sensing and Communications, Ministry of Education, \\University of Electronic Science and Technology of China, Chengdu, P. R. China}
\IEEEauthorblockA{\IEEEauthorrefmark{4}School of Information Technology, Deakin University, Victoria, 3125, Australia}}

\maketitle

\begin{abstract}
Demand response is widely employed by today's data centers to reduce energy consumption in response to the increasing of electricity cost. To incentivize users of data centers participate in the demand response programs, i.e., breaking the ``split incentive'' hurdle, some prior researches propose market-based mechanisms such as dynamic pricing and static monetary rewards. However, these mechanisms are either intrusive or unfair. In this paper, we use time-varying rewards to incentivize users, who have flexible deadlines and are willing to trading performance degradation for monetary rewards, grant time-shifting of their requests. With a game-theoretic framework, we model the game  between a single data center and its users.  Further, we extend our design via integrating it with two other emerging practical demand response strategies: server shutdown and local renewable energy generation. With real-world data traces, we show that a DC with our design can effectively shed its peak electricity load and overall electricity cost without reducing its profit, when comparing it with the current practice where no incentive mechanism is established.
\end{abstract}

% no keywordsg

% \fi

\IEEEpeerreviewmaketitle

\vspace{0.2cm}

\section{Introduction}\label{sec:introduction}
Several Demand Response (DR) strategies for Data Centers (DCs) emerged recently, such as time-shifting of workloads  \cite{Ghatikar2014}. They can effectively reduce/defer a DC's electricity load at specified time in response to the increasing of electricity cost \cite{Wierman2014}. For example, most of today's DCs need to pay both energy and demand charge, where the demand charge is the cost of the DC's peak electricity load during a billing cycle. In this case, the DCs can shed their demand charges via deferring part of the electricity load  from peak hour to off-peak hour \cite{Mathew2014}.

Inevitably, most DR strategies, especially the ones who need workload management, will lead to performance degradation \cite{Liu2014}. For example, time-shifting of workloads can greatly increase the latency of service \cite{Wang2014}. Naturally, users of DCs are not tend to participate in these DR programs if they are not well incentivized \cite{Zhan2015}. However, today's DCs usually charge their users via flat-rate based pricing or usage based pricing. Namely, the cost of a user is decided based on the total usage of DC resources, regardless of when the DC resources are utilized \cite{Li2010}. In this case, users do not share the same incentive as the DCs to participate in specified DR programs, which is referred to as ``split incentive'' \cite{Zhan2016}.

%To break the ``split incentive'' hurdle, several market-based mechanisms emerged recently \cite{Zhan2016,Nasiriani2015,Wang2015,Zhan2015,Wang2014a}. Some of them try to propagate the electricity cost onto the users' costs \cite{Wang2015,Zhan2015} or/and fairly distribute a DC's electricity cost among its users \cite{Nasiriani2015,Wang2014a}. However, in that case, all users are effected regardless of whether they are interested in participating in the DR programs, since every user's cost of DC services may be changed if it was previously charged by a commonly-applied pricing policy, e.g., usage based pricing. Note, we assume that every user was previously charged via a commonly-applied pricing policy and the corresponding cost is referred to as \emph{baseline cost} in the rest of the paper. Meanwhile, the pricing policies employed by DCs to charge their users in these work are always complicated, which may be too costly to implement or prone to be manipulated by DCs \cite{Shakkottai2008}. To overcome these shortcoming, \cite{Zhan2016} offers users of DCs, who are willing to participate in the DR programs, static monetary rewards regardless of whether their workloads are modified for DR.

In this paper, we aim at designing a market-based mechanism for a single DC to break the ``split incentive'' hurdle, which holds the following characteristics:

\begin{itemize}
\item \emph{Monetary incentive}: It provides users of DC with incentives to participate in specified DR programs.% Namely, users are well incentivized to allow DC to modulate their workloads for electricity cost reduction.

\item \emph{Non-intrusiveness}: The DR programs are fully voluntary and the costs and performances of users, who decide not to participate in the DR programs, are not effected.% For example, if a user is previously charged via usage based pricing and it does not want to join in the DR programs, its cost should also be calculated based on its total usage of DC resources and any of its workloads should not be modified for DR.

\item \emph{Deadline awareness}: All users' requests are processed before their deadlines.

\item \emph{Fair reward system}: Only if one participates in the DR programs can gain monetary reward and the reward is proportional to the user's contribution to the DR programs.

\item \emph{Simplicity and max-cost guarantee}: The pricing policy, which is used to charge users of DC, should be easy to understand. Meanwhile, it can guarantee that a user's cost for the DC services will not exceed the \emph{baseline cost}. 

\end{itemize}
Note, we assume that every user was previously charged via a commonly-applied pricing policy, e.g., usage based pricing \cite{Li2010}, and the corresponding cost is referred to as \emph{baseline cost}.

We use time-varying rewards, on top of the commonly applied pricing policy such as flat-rate based pricing or usage based pricing, to incentivize users of DCs grant time-shifting of their requests for DR. With our design, users, who have tight deadlines and/or are not interested in trading performance degradation for monetary rewards, will be charged by the original pricing policy, and their requests will never be deferred. Namely, it is non-intrusive and deadline aware. Meanwhile, we only reward users, whose requests are deferred, and the reward is proportional to the amounts of requests being deferred. Thus, our reward system is fair. Moreover, our pricing policy and reward system can be easily understand and users' maximum costs with our design will never exceed the \emph{baseline costs}.

%With our design, DC periodically releases the reward for granting time-shifting of requests to its users. After receiving the reward information, users respond to  whether they want to trading time-shifting of their requests for the monetary rewards. Next, DC defers part/all of the deferrable requests and reward the owners of the deferred requests based on the pre-released reward information and the amount of deferred requests during each period of time. Here, the deferrable requests means the requests owned by the users who decide to participate in the DR programs.

With our design, the DC needs to make a suitable trade-off between workload management flexibility and monetary reward. Namely, via giving users higher rewards, the  users are more likely to allow the DC to time-shift their requests. On the contrary, if the rewards are relatively low, the DC has less flexibility to response to the electricity price signals.  
The key contributions of this paper are listed as follows:
\begin{itemize}
\item We propose a non-intrusive and fair reward system to incentivize users  with flexible deadlines grant time-shifting of their requests. In this case, the DC has more workload management flexibility to manage its electricity load and thus reduce its electricity cost. 

\item With a game-theoretic framework, we model the game between the DC and its users. We first study users' dominant strategies for deciding whether to join in the DR programs. Then, DC's optimal choices of the time-varying rewards are deduced. Specifically, the DC aims at minimizing its electricity cost without reducing of its profit and the problem of electricity cost minimization is formulated via a convex program.

\item We extend our design via integrating it with each one of the two emerging practical DR strategies: server shutdown \cite{Zhan2016,Paul2015,Li2015} and local renewable energy generation \cite{Zhan2015,Deng2014,Ghamkhari2013}. Therefore, our design can apply to different combinations of DR strategies.
\end{itemize}

In the rest of the paper, we first outline several prior work about using market-based mechanisms to incentivize users of DCs participate in the DR programs in section \ref{sec:relatedworks}. In section \ref{sec:problemformulation}, user's surplus and DC's profit are formulated. We model the game between users and DC in section \ref{sec:interaction} and extend our design in section \ref{sec:extension}. We evaluate our design with real-world data traces and conclude this paper in section \ref{sec:casestudy} and \ref{sec:conclusion}, respectively.

\section{Related Works}\label{sec:relatedworks}

With specified DR programs, DC can effectively reduce its electricity load \cite{Liu2014}, as well as the demand charge and the overall electricity cost \cite{Mathew2014,Xu2014}. Considering that most DR strategies involve management of users' workloads, e.g., time-shifting of requests, several market-based mechanisms have been proposed to incentivize users grant workload management\cite{Zhan2016,Nasiriani2015,Wang2015,Zhan2015,Wang2014a}. 

\cite{Zhan2015} propagated the overall electricity cost onto the users' costs via charging users by Time-of-Use (ToU) pricing. Namely, at each period of time, the price of an instance of DC, i.e., a combination of DC resources such as CPU, memory and storage, is set to be proportional to the prediction of the DC's electricity load. With this pricing policy, users, who can tolerate performance degradation and are sensitive to price, may reduce their purchases of instances at peak hour in response to the increasing of cost. In this case, the DC's peak electricity load and its demand charge can be reduced. 

\cite{Nasiriani2015} fairly distributed a DC's overall electricity cost among its users. With \cite{Nasiriani2015}'s design, if the DC's electricity cost is mainly decided by its peak electricity load, users of the DC can gain considerable cost reduction via reducing its demand when an overall peak occurs. Thus, users are well incentivized to participate in the DR programs to shed the DC's peak electricity load.

However, both \cite{Zhan2015} and \cite{Nasiriani2015} are intrusive since the costs of all users are effected regardless of whether they are interested in participating in the DR programs. Meanwhile, they do not guarantee that the cost of a user with their designs will never exceed the \emph{baseline cost}. 

\cite{Zhan2016} proposed Usage-based Pricing with Monetary Reward (UPMR). \cite{Zhan2016}'s mechanism is non-intrusive and can provide max-cost guarantee. At the beginning of each billing cycle, DC releases the reward function to its users. Then, each user can make the decision of whether to join the DR programs in the upcoming billing cycle based on the reward function and its sensitivity to the performance degradation. The key difference between \cite{Zhan2016}'s work and our design is that: the monetary reward offered by \cite{Zhan2016} is static, i.e., it will remain unchanged during a billing cycle, and the reward is given to all users, who are willing to participate in the DR programs, regardless of whether their workloads are modified for DR. Thus, \cite{Zhan2016}'s reward system is not fair. On the contrary, In this paper, we provide users with time-varying monetary rewards and users can change their decisions of whether to join the DR programs during a billing cycle based on the real-time reward. In this case, the amount of users, who have joined in the DR programs, is time-varying, and thus the DC embraces more load flexibility. Moreover, the reward given to each user with our design is proportional to the user's contribution to the DR program, i.e., the amount of requests being deferred, which is more fair.

\section{Problem Formulation}\label{sec:problemformulation}
A time-slotted system is used in this paper. Namely, we divide a billing cycle into $\tau$ time slots, with equally length of $T$ (hour). In this section, we formulate a user's surplus, as well as a DC's profit.

\subsection{User's Surplus}\label{subsec:userssurplus}
Let $V_{i}[t]$ denote the net utility can be gained by user $i$ after processing one of its request generated at time slot $t$ without participating in any DR programs, where $i \in \{1,\cdots,I\}$ denotes the index for users and $t \in \{1,\cdots,\tau\}$ denotes the index for time slots. Here, $I$ denotes the number of active users. Let $\delta_{i}[t]$ ($\$$) denote the average \emph{baseline cost} for processing one request generated by user $i$ at time slot $t$. Further, let $\kappa_{i}[t]$ and $\gamma[t]$ denote the net utility loss and financial reward caused by modifying one of user $i$'s request generated at time slot $t$, respectively. 

We formulate user $i$'s surplus after processing one of its request generated at time slot $t$, which is denoted by $S_{i}[t]$, as below:
\begin{equation}\label{equ:usersurplus}
S_{i}[t]=V_{i}[t]-\delta_{i}[t]+\mathbbm{1}_{i}[t](\gamma[t]-\kappa_{i}[t]),
\end{equation}
where $\mathbbm{1}_{i}[t]$ is a binary variable and indicates whether the request is modified for DR. Namely, if $\mathbbm{1}_{i}[t]=1$, the request is modified, and thus user $i$ will loss $\kappa_{i}[t]$ net utility but enjoy $\gamma[t]$ monetary reward. On the contrary, if $\mathbbm{1}_{i}[t]=0$, the request has not been modified and thus no net utility is lost and no reward is given. 

\subsection{DC's Profit}\label{subsec:dcsprofit}
\textbf{Revenue}:
Suppose that the DC's revenue is constituted by all of its active users' \emph{baseline costs}. Thus, we formulate the DC's revenue over a billing cycle via
\begin{equation}\label{equ:reveune}
\text{Revenue}=\sum_{t=1}^{\tau} \sum_{i=1}^{I} \lambda_{i}[t]\delta_{i}[t],
\end{equation}
where $\lambda_{i}[t]$ denotes the amount of requests generated by user $i$ at time slot $t$.

\textbf{Electricity cost}:
The electricity cost of today's DC is usually constituted by two parts: energy charge and demand charge, where energy charge is decided by the DC's total energy consumption and demand charge is computed based on the DC's peak electricity load. Specifically, the demand charge can weigh heavily on a DC's overall electricity cost \cite{Xu2014,Islam2015}. Let $\alpha[t]$ (\$/KWh) denote the price of energy charge at time slot $t$ and $\beta_{j}$ (\$/KW) denote the price of type-$j$ demand charge. Here, $j \in \{1,\cdots,J\}$ is the index for types of demand charge and $J$ measures the total types of demand charge. For example, in \cite{Mathew2014}, the authors divided the whole billing cycle into on-peak and off-peak periods and the prices of demand charge during on-peak and off-peak periods can be different. Therefore, we formulate the DC's electricity cost over a billing cycle via 
\begin{equation}\label{equ:cost}
\text{Cost}=\sum_{t=1}^{\tau} T\alpha[t]P[t] +\sum_{j=1}^{J} \max_{t \in A_{j}} \beta_{j}P[t],
\end{equation}
where $A_{j}$ denotes the periods of time falling into the type-$j$ demand charge window. Here, $P[t]$ (KW) denotes the average electricity load of the DC at time slot $t$.

Like in \cite{Paul2015}, the average electricity load of the DC at time slot $t$ is formulated as
\begin{equation}\label{equ:pt}
P[t]=E_{pue}[t]\sum_{n=1}^{N}(e_{0}+u_{n}[t]e_{1}),
\end{equation}
where $E_{pue}[t]$ denotes the average ratio of overall DC energy consumption to IT energy usage at time slot $t$, $n \in \{1,\cdots,N\}$ is the index for active servers, $N$ denotes the total amount of active servers, $e_{0}$ (KW) denotes the power of an idle server, $e_{1}$  (KW) measures the increase of power of a server with unit increase in CPU utilization and $u_{n}[t]$ denotes the average CPU utilization of server $n$ at time slot $t$. 

Without loss of generality, we assume that each request needs the same amount of DC resources to be fully processed and each server can handle the same amount of requests within a time slot. Meanwhile, according to \cite{Lin2013}, the optimal workload dispatching rule is to distribute all requests equally on all active servers. Thus, we formulate the average CPU utilization ratio of server $n$ at time slot $t$ via
\begin{equation}\label{equ:utilization}
u_{n}[t]=\frac{\hat{\lambda}[t]}{N\nu},
\end{equation}
where $\hat{\lambda}[t]$ denotes the  aggregate requests scheduled at time slot $t$ and $\nu$ denotes the maximum amount of requests can be processed by a server within a time slot  in fully utilized condition.

We use time-shifting of users' requests in this paper to shed the DC's peak electricity load. Let $D$ denote the maximum length of deferrable periods of time. Namely, all requests generated at time slot $t$ should  be scheduled no later than time slot $t+D$. Let $\eta_{d}[t]$ denote the aggregate requests generated at time slot $t$ and scheduled at time slot $t+d$, where $d \in \{0,\cdots,D\}$ is the index for the amount of time slots of request being deferred. Specifically, $\eta_{0}[t]$ measures the amount of  requests generated at time slot $t$ and scheduled at the same time slot, i.e., they are not deferred. On the contrary, $\sum_{d=1}^{D} \eta_{d}[t]$ measures the amount of deferred requests generated at time slot $t$. Thus, we formulate the aggregate requests scheduled at time slot $t$ via
\begin{equation}\label{equ:aggregatescheduledr}
\hat{\lambda}[t]=\sum_{d=0}^{D} \eta_{d}[t-d].
\end{equation}

\textbf{Reward}:
We give monetary rewards to users for allowing deferring their requests and the rewards are proportional to the amounts of requests being deferred. Thus, we formulate the total amount of rewards given to the users over a billing cycle via
\begin{equation}\label{equ:reward}
\text{Reward}=\sum_{t=1}^{\tau} \gamma[t]\sum_{d=1}^{D} \eta_{d}[t],
\end{equation}

\textbf{Profit}:
We define the profit of the DC over a billing cycle via
\begin{equation}\label{equ:profit}
\text{Profit}=\text{Revenue}-\text{Cost}-\text{Reward}.
\end{equation}
Note, the DR programs have no impact on the Revenue term, since it is fully constituted by the \emph{baseline costs}. Thus, we treat the Revenue term as a given constant in the rest of the paper.
\begin{figure}[!t]
	\centering
	\includegraphics[width=0.8\linewidth]{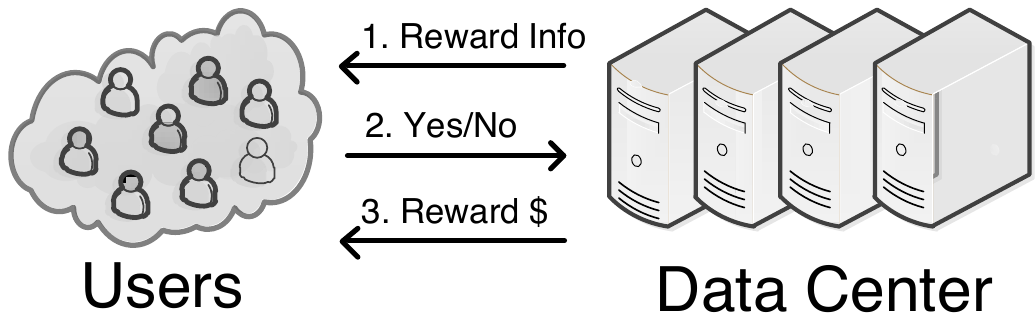}
	\caption{Outline of the interactions between users and DC.}\label{fig:example}
\end{figure}

\section{Game-Theoretic Framework}\label{sec:interaction}
In this paper, we assume that the DC has perfect knowledge of the aggregate requests in the upcoming billing cycle. A game-theoretic framework is proposed for the DC to deduce the optimal choices of rewards in this section.
The interactions between users and DC is shown in Fig. \ref{fig:example}. At the beginning of each time slot, DC first releases the reward information, i.e., $\gamma[t]$, to users. Next, users answer whether they are willing to allow DC to defer their requests generated at this time slot. Finally, DC schedules users' requests and rewards users whose requests are deferred.
% In this section, we model a user's decision of whether to participate in the DR programs, as well as the DC's decisions of reward and  scheduling of requests.

\subsection{Users' Dominant Strategies}\label{subsec:userdecision}
Every user needs to make the decision of whether to allow DC to defer its requests generated at the upcoming time slot based on the reward, i.e., $\gamma[t]$, and its sensitivity to time-shifting of its requests, i.e., $\kappa_{i}[t]$. 
\begin{assumption}\label{ass:rational}
Suppose every user is \emph{rational}, which means that it wants to maximize its surplus as defined in (\ref{equ:usersurplus}). 
\end{assumption}

\begin{assumption}\label{ass:pricetaker}
Suppose that the DC holds an enormous amount of users. In this case, each user's contribution to the DC's overall electricity load is negligible, and thus every user cannot strategically manipulate the price of DC services and the rewards given to users. Namely, the users are \emph{price takers}.
\end{assumption}

\begin{theorem}\label{the:dominant}
From Assumption \ref{ass:rational} and \ref{ass:pricetaker}, $\forall i,t$, if the reward exceeds the net utility loss caused by time-shifting of a user's requests, i.e., $\gamma[t]>\kappa_{i}[t]$, the dominant strategy for the user is to grant time-shifting of its requests. On the contrary, $\forall i,t$, if $\gamma[t] \leq \kappa_{i}[t]$, the dominant strategy for the user is not to participate in the DR programs.
\end{theorem}

%\begin{proof}\label{pro:dominant}

%\end{proof}
The proof of Theorem \ref{proo:dominant} can be found in Appendix \ref{proo:dominant}. 
\subsection{DC's Optimal Decisions}\label{subsec:dcdecision}

\textbf{Variables}:
The DC needs to make two decisions at each time slot: $\gamma[t]$ and $\eta_{d}[t]$. Specifically, $\gamma[t]$ is about the reward given to users for granting time-shifting of their requests and $\eta_{d}[t]$ elaborates the detail of request distribution.

\textbf{Constraints}:
First, we define the domain of the variables:
\begin{equation}\label{con:domainreward}
Lb[t] \leq \gamma[t]\leq Ub[t], \quad \forall t,
\end{equation}
\begin{equation}\label{con:domaineta}
\eta_{d}[t] \geq 0, \quad \forall d,t,
\end{equation}
where  $Lb[t]$ and $Ub[t]$ denote the lower bound and upper bound of net utility loss factor, i.e., $\kappa_{i}[t]$, of elastic users, respectively.
Here, the elastic users are the ones  who have flexible deadlines and are will allow DC to time-shift their requests if they are well incentivized. On the contrary, the users, who have tight deadlines or/and are not interested in trading performance degradation for monetary rewards, are referred to as the inelastic users. Specifically, for all inelastic users, we let their net utility loss factor $\kappa_{i}[t]$ to be infinity.
In this case, $\forall \gamma[t] \in [0,Lb[t]]$, there is no user will join the DR programs. Meanwhile, $\forall \gamma[t] \in [Ub[t],\infty)$, all elastic users are willing to allow DC to time-shift their requests. Thus, in (\ref{con:domainreward}), we indicate that the DC has no interest to set the reward lower than $Lb[t]$ or higher than $Ub[t]$.

Second, the amount of requests scheduled at each time slot should not exceed the capacity of the DC, which is formulated as
\begin{equation}\label{con:capacity}
\sum_{d=0}^{D} \eta_{d}[t-d]\leq N\nu.
\end{equation}
With (\ref{con:capacity}), we ensure that the average CPU utilization ratio of each server  of the DC defined by (\ref{equ:utilization}) will never exceed 1.

Third, all requests generated at each time slot should  be scheduled before the deadline, which is formulated as
\begin{equation}\label{con:aggregaterequest}
\sum_{d=0}^{D} \eta_{d}[t] = \lambda[t], \quad \forall t,
\end{equation}
where $\lambda[t]$ denotes the aggregate requests of users generated at time slot $t$ and
\begin{equation}\label{equ:aggregategeneraterequest}
\lambda[t]=\sum_{i=1}^{I} \lambda_{i}[t], \quad \forall t.
\end{equation}

Fourth, we notice that the uniform distribution is widely used in economic models \cite{Gibbens2000}. Thus, we assume that, $\forall t$, the net utility loss factors $\kappa_{i}[t]$ of elastic users are uniformly distributed from $Lb[t]$ to $Ub[t]$. Moreover, without loss of generality, suppose that each user will generate the same amount of requests at each time slot, i.e., $\forall t$ and $\forall i^{1},i^{2}\in \{1,\cdots,I\}$, $\lambda_{i^{1}}[t]=\lambda_{i^{2}}[t]$. 
From Theorem \ref{the:dominant}, we obtain that, $\forall t$, the amount of deferrable requests, i.e., the sum of requests generated by users who grant time-shifting of their requests, equals $\pi[t] \lambda[t]\frac{\gamma[t]-Lb[t]}{Ub[t]-Lb[t]}$, where $\pi[t]$ denotes the ratio of the elastic users to the overall users at time slot $t$.   Obviously, the amount of requests being deferred should be no higher than the sum of deferrable requests, which is modeled as
\begin{equation}\label{con:alldeferredrequests}
\sum_{d=1}^{D} \eta_{d}[t] \leq \pi[t] \lambda[t]\frac{\gamma[t]-Lb[t]}{Ub[t]-Lb[t]}, \quad \forall t.
\end{equation}

Last but not least, in this paper, we aim at minimizing the electricity cost of the DC while ensuring that the profit with our design is no less than the one without any DR programs and incentive mechanisms, which is modeled as
\begin{equation}\label{con:profitneutrality1}
\text{Revenue}-\text{Cost}-\text{Reward} \geq \text{Revenue}-\text{Baseline Cost},
\end{equation}
where the Revenue, Cost and Reward terms are defined by (\ref{equ:reveune}), (\ref{equ:cost}) and (\ref{equ:reward}), respectively. Here, the Baseline Cost term is the electricity cost of the DC if no DR strategy is implemented. Correspondingly, there is also no reward given to the users. Further, we simplify (\ref{con:profitneutrality1}) and reformulate it as
\begin{equation}\label{con:profitneutrality}
\text{Cost}+\text{Reward} \leq \text{Baseline Cost}.
\end{equation}

\textbf{Minimization of electricity cost}:
We define the problem of minimization of the DC's electricity cost without reducing its profit via
\begin{equation}\label{problem:dc}
\begin{aligned}
& \underset{\gamma[t],\eta_{d}[t]}{\text{min}}
& &\text{Cost}\\
& \ \text{s.t.}
& & (\ref{con:domainreward})-(\ref{con:aggregaterequest}), (\ref{con:alldeferredrequests}), (\ref{con:profitneutrality}).\\
\end{aligned}\!\!\!
\end{equation}
where the Cost term is defined in (\ref{equ:cost}). 

Apparently, problem (\ref{problem:dc}) is not convex caused by the non-convex term Reward (\ref{equ:reward}). 

\begin{theorem}\label{the:optimalreward}
For any given $\eta_{d}[t]$, $\gamma^{*}[t]$, which is defined as
\begin{equation}\label{con:alldeferredrequests1}
\gamma^{*}[t]=\frac{(Ub[t]-Lb[t])\sum_{d=1}^{D} \eta_{d}[t]}{\pi[t] \lambda[t]} +Lb[t], \quad \forall t,
\end{equation}
is one of the optimal choice of $\gamma[t]$ of problem (\ref{problem:dc}).
\end{theorem}
%\begin{proof}\label{proof:optimalreward}

%\end{proof}

The proof of Theorem \ref{the:optimalreward} can be found in Appendix \ref{proo:optimal}.
From Theorem \ref{the:optimalreward}, we let $\gamma[t]=\gamma^{*}[t]$, $\forall t$. 
Further, we update the problem (\ref{problem:dc}) to be 
\begin{equation}\label{problem:dcupdate}
\begin{aligned}
& \underset{\eta_{d}[t]}{\text{min}}
& &\text{Cost}\\
& \ \text{s.t.}
& & (\ref{con:domaineta})-(\ref{con:aggregaterequest}), (\ref{con:profitneutrality}).\\
\end{aligned}\!\!\!
\end{equation}
Apparently, problem (\ref{problem:dcupdate})  is convex, which can be effectively solved via a convex tool such as CVX \cite{CVX}.

\section{Extensions}\label{sec:extension}
We extend our design via integrating workload time-shifting with another emerging practical DR strategy: server shutdown \cite{Zhan2016,Paul2015,Li2015} or local renewable energy generation \cite{Zhan2015,Deng2014,Ghamkhari2013} in this section.

\subsection{Server Shutdown}\label{subsec:servershutdown}
Considering that the proportionality degree of today's commercial server is relatively low, e.g., an idle server can consume over $85\%$ of peak power \cite{Qureshi2010}, shutting down the idle servers can save a huge amount of energy.
In this paper, we take into consideration the energy overhead and the wear-and-tear cost caused by turning on/off servers as in \cite{Zhan2016}. In this case, at each time  slot, the DC also needs to decide how many servers should be switched on/off, which is denoted by $m_{on}[t]$ and $m_{off}[t]$, respectively.

As in \cite{Zhan2016}, we define the  electricity cost of the DC over a billing cycle with server shutdown, which is denoted by $\text{Cost}_{S}$, as
\begin{equation}\label{equ:costservershutdown}
\begin{aligned}
\text{Cost}_{S}=&\sum_{t=1}^{\tau} \alpha[t](TP_{s}[t]+E_{pue}P_{o}[t]) +\\
&\sum_{j=1}^{J} \beta_{j} \max_{t \in A_{j}} \left(P_{s}[t]+\frac{E_{pue}P_{o}[t]}{T}\right),
\end{aligned}
\end{equation}
where $P_{s}[t]$ and $P_{o}[t]$ denote the average electricity load of the DC with server shutdown and the average energy overhead of turning on/off servers at time slot $t$, respectively. Here, both of the definitions of $P_{s}[t]$ and $P_{o}[t]$ can be found in  \cite{Zhan2016}. Moreover, the wear-and-tear cost of servers over a billing cycle, which is denoted by Wear, is also defined as in  \cite{Zhan2016}.

%. We formulate  $P_{o}[t]$ by
%\begin{equation}\label{equ:po}
%P_{o}[t]=o_{on}m_{on}[t]+o_{off}m_{off}[t],
%\end{equation}
%where $o_{on}$ (KWh) and $o_{off}$ (KWh) measure the energy consumption of turning on/off a server, respectively.
%
%Next, as in \cite{Zhan2016}, we formulate the Wear term as
%\begin{equation}\label{equ:wear}
%\text{Wear}=\sum_{t=1}^{\tau}w_{on}m_{on}[t]+\sum_{t=1}^{\tau}w_{off}m_{off}[t],
%\end{equation}
%where $w_{on}$ ($\$$) and $w_{off}$ ($\$$) measure the wear-and-tear cost of servers due to turning on/off servers, respectively.

The problem of minimizing the DC's electricity cost with server shutdown without reducing its profit is modeled as
\begin{equation}\label{problem:dcservershutdown}
\begin{aligned}
& \underset{\begin{subarray}{c}\eta_{d}[t],\\ m_{on}[t],\\m_{off}[t] \end{subarray}}{\text{min}}
& &\text{Cost}_{S}\\
& \ \text{s.t.}
& & (\ref{con:domaineta}), (\ref{con:aggregaterequest}), \\
&&& \text{Cost}_{S}+\text{Reward}+\text{Wear}\leq \text{Baseline Cost},\\
&&& m_{on}[t] \geq 0, \quad m_{off}[t] \geq 0, & \forall t,\\
&&& \frac{\hat{\lambda}[t]}{\nu} \leq m[0]+\sum_{t^{'}=1}^{t} (m_{on}[t]-m_{off}[t]) \leq N, & \forall t,\\
\end{aligned}\!\!\!
\end{equation}
where $m[0]$ denotes the amount of switched on servers at the beginning of the billing cycle.
Here, the last constraint of problem (\ref{problem:dcservershutdown}) indicates that the amount of switched on servers at each time slot should be large enough to handle all scheduled requests, but no larger than the quantity of available servers, which is denoted by $N$. Like problem (\ref{problem:dcupdate}), problem (\ref{problem:dcservershutdown}) is also convex.

\subsection{Local Renewable Energy Generation}\label{subsec:renewable}
Recently, many state-of-the-art DCs have invested heavily to develop local renewable energy generation, such as Apple and Google \cite{Liu2014}. Let $G[t]$ (KWh) denote the renewable energy generated at time slot $t$. As in \cite{Zhan2015}, we assume that the information of renewable energy can be perfectly predicted by the DC and the part of  renewable energy, which is not used by the DC, is dropped. Thus, the electricity cost with local renewable energy generation, which is referred to as Cost$_{R}$, is defined by
\begin{equation}\label{equ:costrenewable}
\begin{aligned}
\text{Cost}_{R}=&\sum_{t=1}^{\tau} \alpha[t]\left[TP[t]-G[t]\right]^{+} +\\
&\sum_{j=1}^{J} \beta_{j} \max_{t \in A_{j}} \left[P[t]-\frac{G[t]}{T}\right]^{+},
\end{aligned}
\end{equation}
where $[x]^{+}$ represents the function $\max\{x,0\}$.

Next, we model the problem of minimization of the DC's electricity cost with local renewable energy generation without reducing its profit via
\begin{equation}\label{problem:dcrenewable}
\begin{aligned}
& \underset{\eta_{d}[t]}{\text{min}}
& &\text{Cost}_{R}\\
& \ \text{s.t.}
& & (\ref{con:domaineta})-(\ref{con:aggregaterequest}), \\
&&& \text{Cost}_{R}+\text{Reward} \leq \text{Baseline Cost},\\
\end{aligned}\!\!\!
\end{equation}
which is also convex.

\section{Case Studies}\label{sec:casestudy}
Three real-world data traces are used in this paper to represent users' requests: Gmail U.S., Gmaps U.S. and Youtube U.S. from January 1, 2014 to January 31, 2014 \cite{Googledata}. According to the electric rates of industrial power service offered by South Carolina Electric and Gas  \cite{SCEG}, $\forall t$, let $\alpha[t]=\$0.05207$ per KWh, $J=1$ and $\beta_{1}=\$15.59$ per KW. Let $T=1$ (hour) and $\tau=720$. Namely, the billing cycle is 30 days basis. Let $e_{0}=0.1$ KW, $e_{1}=0.1$ KW, $\nu=20$ and $E_{pue}=1.2$. $\forall t$, let $\pi[t]=0.5$, $Lb[t]=10e^{-4}$ ($\$$) and  $Lb[t]=10e^{-3}$ ($\$$). Next, in this section, we first evaluate our work by comparing its performance with the one without any incentive mechanisms, which represents the current practice of DC and is referred to as \textbf{Baseline}. Then, we study the impact of integrating our design with another DR strategy on the reduction of peak electricity load and overall electricity cost.

\subsection{Performance Evaluation}\label{subsec:performanceevaluation}

\begin{figure}[!t]
\centering
\subfigure[]
{ \label{fig:performance_peak}
\includegraphics[width=0.465\columnwidth]{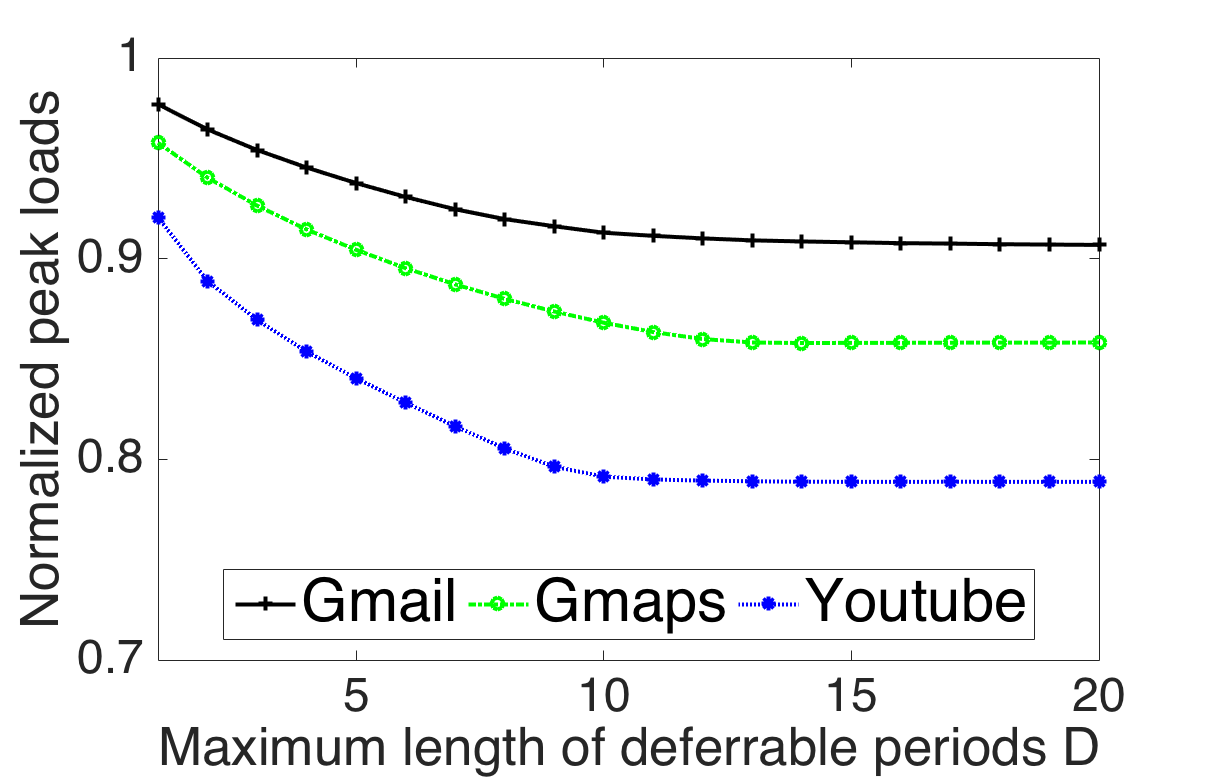}
}
\subfigure[] 
{ \label{fig:performance_cost}
\includegraphics[width=0.465\columnwidth]{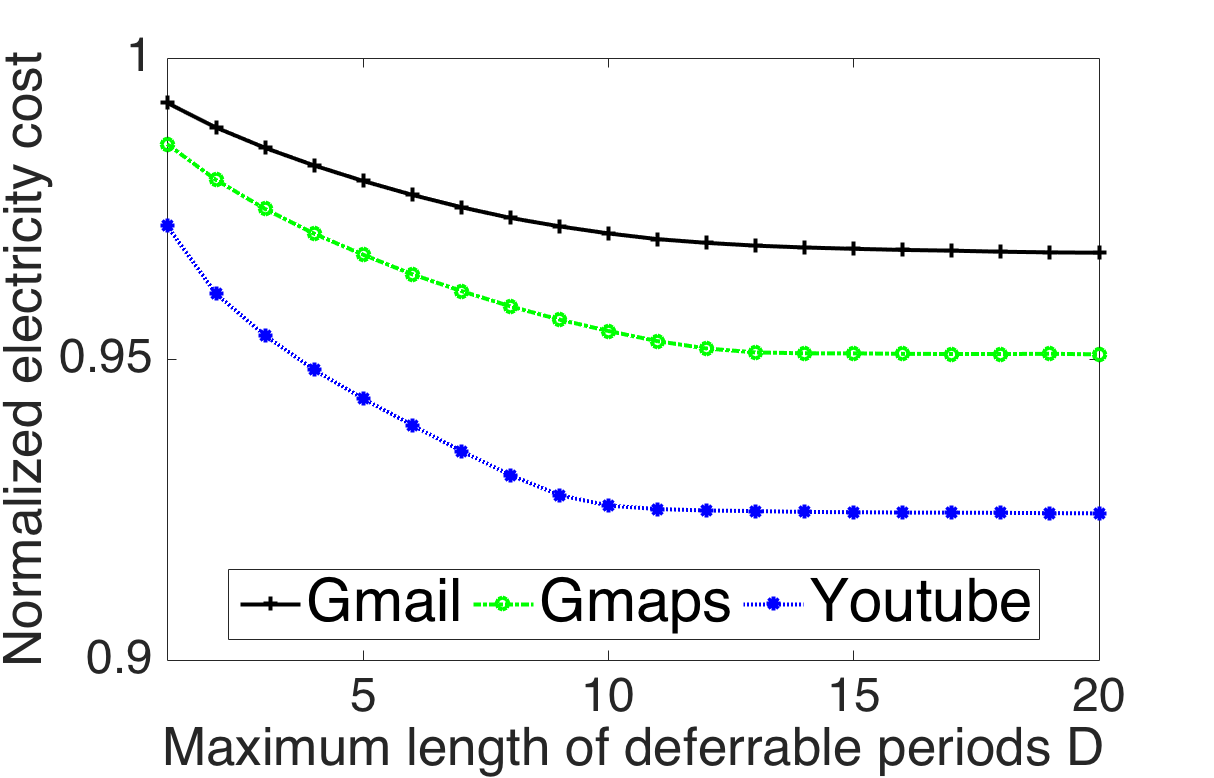}
}
\centering\caption{Normalized results of a DC with our design: (a) Peak electricity load, (b) Electricity cost.}\label{fig:performance}
\end{figure}

We study the DC's peak electricity load and electricity cost with our design under different maximum lengths of deferrable periods of time, i.e., $D$. Fig. \ref{fig:performance} shows the normalized results with respect to the results of Baseline.

Three information can be found from Fig. \ref{fig:performance}: 1) With the increasing of the maximum length of deferrable periods of time, our design can shed the DC's peak electricity load and reduce its overall electricity cost more effectively. 2) When the maximum length of deferrable periods is relatively large, e.g., larger than 10 time slots as shown in Fig. \ref{fig:performance}, the effectiveness of our design cannot be greatly increased via further increasing the maximum length of deferrable periods of time. 3) DC with our design always outperforms the one without any DR  programs and incentive mechanisms. For example, when $D=10$, the DC with data traces of Youtube U.S. can shed its peak electricity load by $20.9 \%$ and reduce its overall electricity cost by $7.4\%$.

\subsection{Impact of Server Shutdown and Local Renewable Energy Generation}\label{subsec:impact}
\begin{figure}[!t]
\centering
\subfigure[]
{ \label{fig:impact_turn_peak}
\includegraphics[width=0.465\columnwidth]{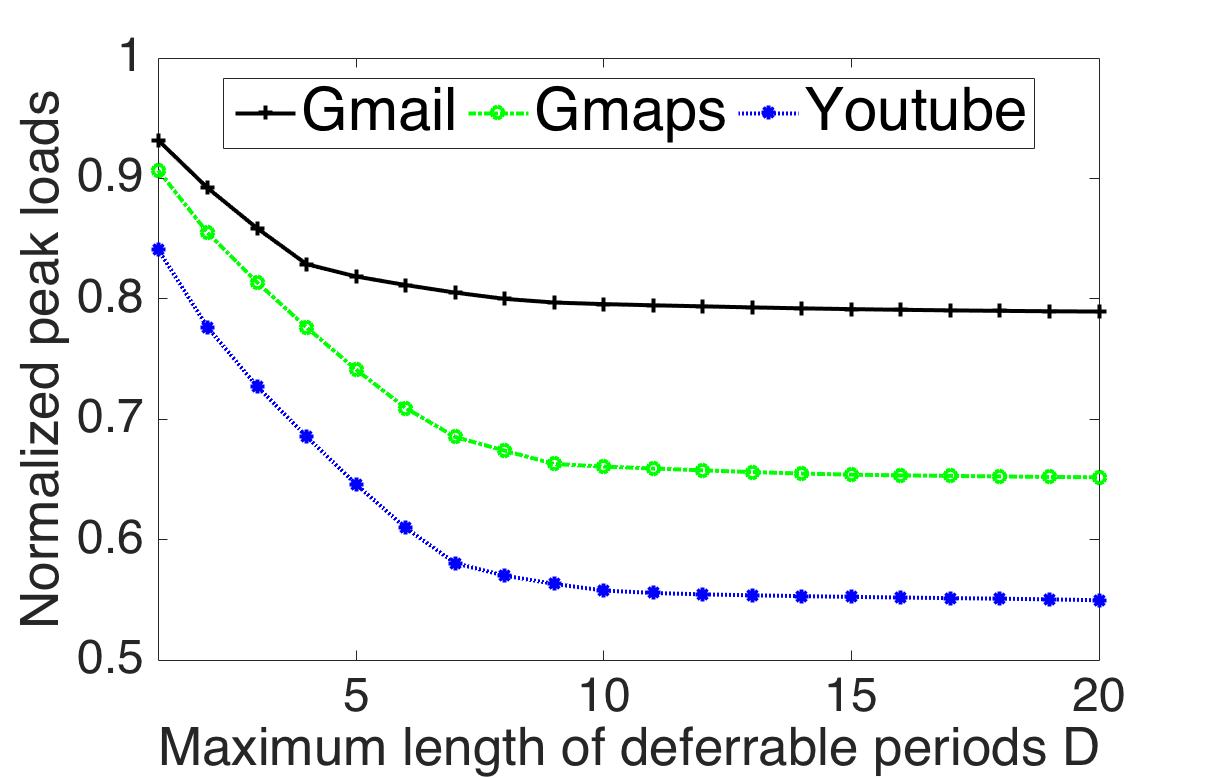}
}
\subfigure[] 
{ \label{fig:impact_turn_cost}
\includegraphics[width=0.465\columnwidth]{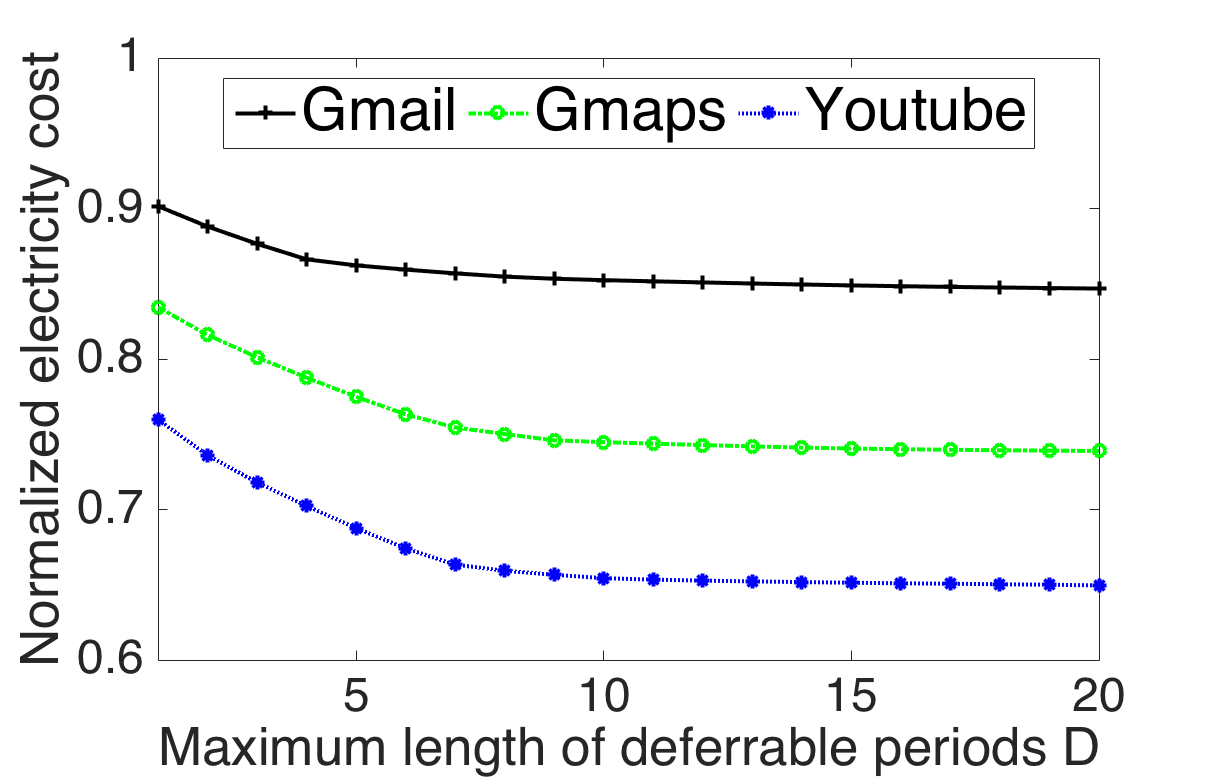}
}
\centering\caption{Normalized results of a DC with our design and server shutdown: (a) Peak electricity load, (b) Electricity cost.}\label{fig:impact_turn}
\end{figure}

\begin{figure}[!t]
\centering
\subfigure[]
{ \label{fig:impact_green_peak}
\includegraphics[width=0.465\columnwidth]{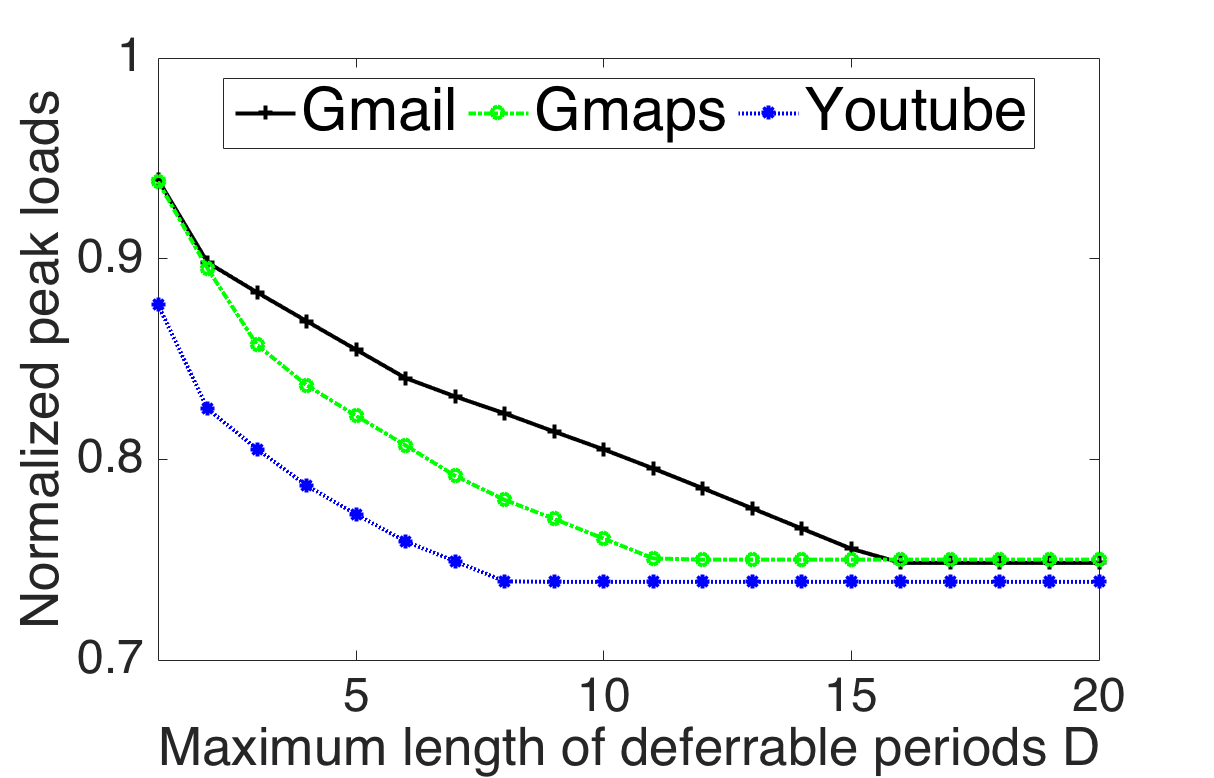}
}
\subfigure[] 
{ \label{fig:impact_green_cost}
\includegraphics[width=0.465\columnwidth]{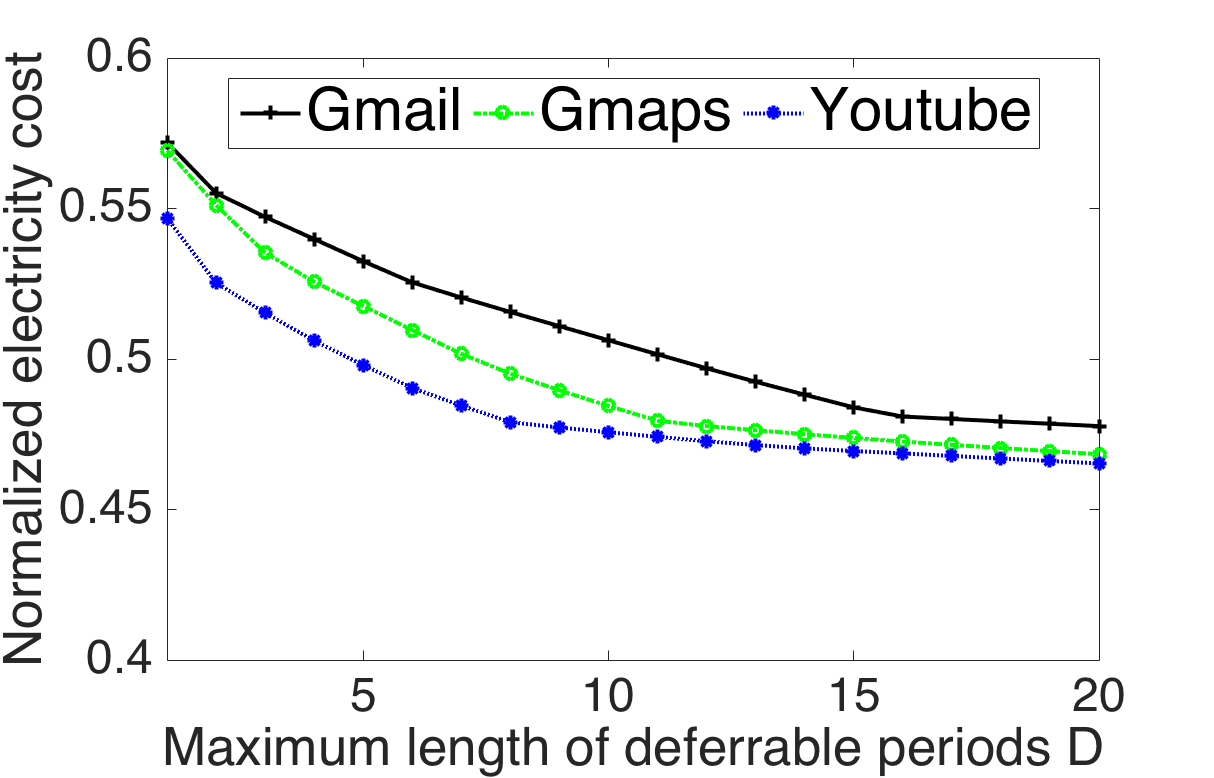}
}
\centering\caption{Normalized results of a DC with our design and local renewable energy generation: (a) Peak electricity load, (b) Electricity cost.}\label{fig:impact_green}
\end{figure}

We integrate our design with another emerging practical DR strategy: server shutdown or local renewable energy generation, and evaluate their performances. Fig. \ref{fig:impact_turn} and Fig.
\ref{fig:impact_green} show the normalized results of a DC with server shutdown and local renewable energy generation, respectively. Here, the base for both normalizations is the results of Baseline. Specifically, the DC as shown in Fig.
\ref{fig:impact_green} is equipped with several wind turbines and the wind speed data is gathered from January 1, 2012 to January 30, 2012 and is available in \cite{winddata}. By analyzing Fig. \ref{fig:impact_turn} and
\ref{fig:impact_green}, the three information found in section \ref{subsec:performanceevaluation} still hold.  Moreover, By comparing Fig. \ref{fig:impact_turn} and 
\ref{fig:impact_green} against Fig. \ref{fig:performance}, we find that the peak electricity load and the overall electricity cost of a DC with our design can be further reduced via integrating with other DR strategies such as server shutdown and local renewable energy generation.

\section{Conclusion}\label{sec:conclusion}
In this paper, we proposed time-varying rewards to incentivize users of DCs grant time-shifting of their requests. Our design is non-intrusive, deadline aware and easy to understand. Moreover, our reward system is fair and we can provide users with max-cost guarantee. With a game-theoretic framework, we modeled the game between users and a single DC and formulated/solved the  problem of minimization of the DC's electricity cost without reducing its profit. With real-world data traces, we showed that our design can greatly reduce the DC's peak electricity load and overall electricity cost. We also extended our design to another emerging practical scenario where the DC has employed workload time-shifting and server shutdown or local renewable energy generation simultaneously. Accordingly, our design can apply to a DC with different combinations of DR strategies.

\appendices

\section{Proof of Theorem \ref{the:dominant}}\label{proo:dominant}
Apparently, $\forall i, t$, from (\ref{equ:usersurplus}), if a user does not grant time-shifting of its requests, its surplus at time time slot $t$, which is referred to as $S_{i}^{No}[t]$, equals 
\begin{equation}\label{equ:usersurplusno}
S_{i}^{No}[t]=\lambda_{i}[t](V_{i}[t]-\delta_{i}[t]).
\end{equation}
On the contrary, if the user decides to participate in the DR programs, its surplus at time slot $t$, which is referred to as $S_{i}^{Yes}[t]$, is
\begin{equation}\label{equ:usersurplusyes}
S_{i}^{Yes}[t]=\lambda_{i}[t](V_{i}[t]-\delta_{i}[t])+\hat{\lambda}_{i}[t](\gamma[t]-\kappa_{i}[t]),
\end{equation}
where $\hat{\lambda}_{i}[t]$ denotes amount of deferred requests generated by user $i$ at time slot $t$. Note, since DC makes the decision of whether to defer part or all of the users requests if the user grant time-shifting of its requests, the user cannot perfectly predict how many requests will be deferred but can ensure that $0\leq \hat{\lambda}_{i}[t]\leq \lambda_{i}[t]$.

\noindent
Next, we prove that Theorem \ref{the:dominant} is correct in two complementary cases: First, $\forall i,t$, if $\gamma[t]>\kappa_{i}[t]$, $S_{i}^{No}[t] \leq S_{i}^{Yes}[t]$. Namely, in this case, the dominant strategy for the user is to join in the DR programs. Second, $\forall i,t$, if  $\gamma[t]\leq \kappa_{i}[t]$, $S_{i}^{No}[t] \geq S_{i}^{Yes}[t]$, and thus the optimal choice for the user is not to participate in the DR program in this case. $\hfill \blacksquare$

\section{Proof of Theorem \ref{the:optimalreward}}\label{proo:optimal}
First, we can easily obtain that $\forall t$, $\gamma^{*}[t] \geq Lb[t]$. Second, from (\ref{con:alldeferredrequests}), we have $\forall t$,
\begin{equation*}
\begin{aligned}
\sum_{d=1}^{D}\eta_{d}[t] &\leq \pi[t] \lambda[t] &\Longrightarrow \\
 \frac{(Ub[t]-Lb[t])\sum_{d=1}^{D} \eta_{d}[t]}{\pi[t] \lambda[t]} &\leq Ub[t]-Lb[t] &\Longrightarrow \\
 \gamma^{*}[t] &\leq Ub[t].
\end{aligned}
\end{equation*}
Meanwhile, $\gamma^{*}[t]$ satisfy constraint (\ref{con:alldeferredrequests}) of problem (\ref{problem:dc}). Thus,  $\gamma^{*}[t]$ is a feasible solution of $\gamma[t]$ of problem (\ref{problem:dc}). Next, we prove Theorem \ref{the:optimalreward} by contradiction. Assume that $\gamma^{*}[t]$ is not an optimal solution of $\gamma[t]$ of problem (\ref{problem:dc}) and let $\gamma^{c}[t]$ denote the true optimal solution of $\gamma[t]$ of problem (\ref{problem:dc}). From (\ref{equ:cost}), $\text{Cost}\arrowvert_{\gamma[t]=\gamma^{c}[t]}=\text{Cost}\arrowvert_{\gamma[t]=\gamma^{*}[t]}$, which contradicts the assumption that $\gamma^{*}[t]$ is not an optimal solution of $\gamma[t]$ of problem (\ref{problem:dc}).
$\hfill \blacksquare$

\end{document}